\documentclass[]{aa}

\usepackage{psfig}

\def\etal{{et\,al. }}
\def\msun{M$_{\odot}$}

\def\degs{\ifmmode ^{\circ}\else$^{\circ}$\fi}
\def\amin{\ifmmode ^{\prime}\else$^{\prime}$\fi}
\def\asec{\ifmmode ^{\prime\prime}\else$^{\prime\prime}$\fi}
\newbox\grsign \setbox\grsign=\hbox{$>$}
\newdimen\grdimen \grdimen=\ht\grsign
\newbox\laxbox \newbox\gaxbox
\setbox\gaxbox=\hbox{\raise.5ex\hbox{$>$}\llap
     {\lower.5ex\hbox{$\sim$}}}\ht1=\grdimen\dp1=0pt
\setbox\laxbox=\hbox{\raise.5ex\hbox{$<$}\llap
     {\lower.5ex\hbox{$\sim$}}}\ht2=\grdimen\dp2=0pt

\def\lax{\mathrel{\copy\laxbox}}

\def\grs{GRS 1915+105}
\def\gro{GRO J1655-40}
\def\1e17{1E 1740.7-2942}
\def\gr17{GRS 1758-258}
\def\4u{4U~1630-47}
\def\cyg{Cyg X-1}\def\gx{GX 339-4}

\begin{document}

   \title{Comptonization and reflection of X-ray radiation and the
     X-ray-radio correlation in the $\chi$-states of GRS  1915+105
     \thanks{complete Table 2 is only available in electronic form at the CDS
     via anonymous ftp to cdsarc.u-strasbg.fr (130.79.128.5) or via http://cdsweb.u-strasbg.fr/cgi-bin/qcat?J/A+A/(vol)/(page)}}

   \titlerunning{Comptonization and Reflection in GRS 1915+105}

   \author{A. Rau\inst{1,2} \& J. Greiner\inst{1,2}}

   \offprints{A. Rau, arau@mpe.mpg.de}

   \institute{Astrophysical Institute
             Potsdam, An der Sternwarte 16, 14482 Potsdam, Germany 
	\and
	     Max-Planck-Institute for extraterrestrial Physics, Giessenbachstrasse, 85748 Garching, Germany}

   \date{Received 8 March 2002; accepted ????}

   \abstract{We present a comprehensive X-ray study of four years of
   pointed {\it RXTE} observations of \grs\ in the $\chi$-state. We
   interpret the behavior of the hard power law tail spectrum as
   coming from inverse Compton scattering of soft disk photons on a
   thermally dominated hybrid corona above the accretion disk. \grs\
   shows a strong, variable reflection amplitude. As in other BHC and
   in Seyfert galaxies, a correlation between the power law slope and
   the reflection was found. Also, the radio fluxes at 2.25\,GHz and
   15\,GHz correlate with the power law slope, thus revealing a
   connection between the outflowing matter and the comptonizing
   region in the $\chi$-states.
   	}

    \maketitle
    
\section{Introduction}

With the discovery of superluminal ejections (Mirabel \& Rodriguez
1994) from the galactic transient source \grs\ and \gro, much
observational and theoretical attention has been directed to the field
of microquasars. Microquasars are thought to be downscaled analogs to
quasars exhibiting much smaller time scales and are therefore
potential laboratories for studying accretion and relativistic jets
near black holes (Mirabel \etal 1992). The most prominent object of
this type is the galactic X-ray binary system \grs\ which shows
dramatic variability (Greiner \etal 1996) in light curve,
quasi-periodic oscillations, phase lags and coherence behavior (Morgan
\etal\ 1997, Muno \etal 2001). \grs\ is the most energetic object
known in our galaxy with a luminosity of $\sim$10$^{38}$ erg/s in the
low state.

\grs\ harbors a black hole of 14\,\msun\ (Greiner \etal 2001b) making
it the most massive stellar black hole known. Being in the galactic
plane and at a distance of $\sim$12~kpc the source suffers a very high
extinction in the optical band of the order of 25-30\,mag (Greiner
\etal 1994, Chaty \etal 1996). Greiner \etal\ (2001a) found the donor
to be an K-M III late type giant as determined from absorption line
measurements in the near infrared.

Black hole transients generally exhibit different states of intensity
and spectrum: the high/soft state with a prominent disk component and
a weak or negligible steep power law tail, the low/hard state with
negligible accretion disk and flat power law. In the intermediate and
very high state both the accretion disk and a power law are seen.

Originally discovered by {\it Granat} (Castro-Tirado \etal 1992), \grs\ was extensively monitored by the {\it RXTE} since 1996 and a number of investigations of these data have been published (e. g. Morgan \etal 1997, Belloni \etal 1997 \& 2000, Muno \etal 1999 \& 2000)

The source evades simple classification, although it seems to spend
most of the time in the very high state. Several attempts to
categorize the behavior of \grs\ have been made in the past. Belloni
\etal (2000) defined 12 different states depending on light curve
variation and hardness colors. One of these states, the so-called
$\chi$-state is characterized by a lack of obvious variations in the
light curve and spectrum and is associated with continuous radio
emission of differing strength. $\chi$-states correspond to the
low/hard state of \grs, exhibiting relatively low flux from the
accretion disk and showing a hard power law tail. Theses states
resemble long variants of the lulls in $\beta$-states, when part of
the inner accretion disk is proposed to be absent. Depending on the
strength of the radio emission, differing phase lag and power spectrum
behavior are seen. During the so-called radio quiet $\chi$-states the
phase lag of the hard X-ray photons to the soft X-ray photons is
partly negative, whereas it is always positive when the radio emission
is strong (Muno \etal 2001). Also, the frequency of the 0.5--10\,Hz
QPO decreases with increasing radio flux.
 
 In this paper we present a comprehensive study of the X-ray spectral
 behavior of \grs\ in the $\chi$-state with over nearly four years of
 observation with {\it RXTE}. The variation of the spectral properties
 is then compared to those of the radio emission. Valuable information
 about the disk geometry and properties of the compact object is
 derived.

\section{Data selection and analysis}

\subsection{{\it RXTE}}

For our investigation of \grs\ we used public {\it RXTE} data from
November 1996 to September 2000 provided by HEASARC. We selected
Proportional Counter Array (PCA) and High Energy X-Ray Timing
Experiment (HEXTE) data of 139 $\chi$-state observations from 89
different days. The selection of the datasets was based on the
$\chi$-states defined by Belloni \etal (2000) and on PCA light curves
provided by
E. H. Morgan\footnote{http://xte.mit.edu/$\sim$ehm/1915\_frames.html}. Data
with more than 30\,min away from the South Atlantic Anomaly and
without X1908+075 in the HEXTE
background\footnote{http://mamacass.ucsd.edu:8080/cgi-bin/HEXTErock.html}
were selected. The lack of obvious variability in the light curve and
its spectral hardness allows long continuous exposure times and
therefore high signal-to-noise ratios. Typically, the spectra during
this state show blackbody emission arising from an accretion disk and
a dominating power law hard energy tail.

We reduced the {\it RXTE} data using the standard reduction script REX
included in the HEAsoft5.04 package. We restricted the analysis to
Standard 2 binned data of all layers of PCU0 of the PCA from
3--25\,keV and HEXTE cluster 0 from 20--190\,keV only. We used PCARSP
7.10 to produce a particular response matrix for each PCU0 dataset. In
order to account for part of the uncertainties in the PCU0 instrument
we added a systematic error of 1\,\% as recommended
(Remillard\footnote{http://lheawww.gsfc.nasa.gov/users/keith/ronr.txt}).

The X-ray spectral fitting was done using XSPEC 11.0 (Arnaud 1996). A
consistent model should fit all {\it RXTE} $\chi$-state
data. Therefore we tested several X-ray radiation models and finally
selected a model consisting of (i) photoelectric absorption (WABS;
Balucinska-Church \& McGammon 1992), (ii) a spectrum from an accretion
disk consisting of multiple blackbody components (DISKBB) and (iii) a
power law spectrum reflected from an ionized relativistic accretion
disk (REFSCH; Fabian \etal 1989, Magdziarz \& Zdziarski 1995).
Several attempts in the past to fit \grs\ spectra have shown
complicated residuals in the soft X-ray band suggesting the existence
of emission and/or absorption features near 6.4\,keV, the energy of
the Fe K$\alpha$ line (Kotani \etal 2000). Because of the low spectral
resolution of $\sim$1\,keV of the PCA at this energy an additional
line fit gives no meaningful results. Therefore we ignored all energy
bins from 4.5 to 8.5\,keV while fitting our model. We fixed the
hydrogen column density at $N_H$=5$\cdot$10$^{22}$\,cm$^{-2}$ as
determined by Greiner \etal (1994) with {\it ROSAT}. It has been shown
that the disk blackbody + power law assumption strongly overestimates
the flux at lower energies compared to the thermal and non-thermal
comptonization models used by Vilhu \etal (2001) and Zdziarski \etal
(2001), respectively. This explains the smaller $N_H$
(2--3$\cdot$10$^{22}$\,cm$^{-2}$) values they found for $\chi$-state
observations. However, because we ignored the energy bins from 4.5 to
8.5\,keV the amount of data bins needed to adjust the hydrogen column
density was to small to let $N_H$ be a free parameter.

We fixed 9 of the combined 16 model parameters
(Table~\ref{fixedpara}), leaving free the accretion disk temperature,
$T_{bb}$, and relative normalization, $K_{bb}\propto R_{in}^2$, the
power law photon index, $\Gamma$, with $N\propto$E$^{-\Gamma}$ and
relative normalization, $K_{po}$ (photons/keV/cm$^2$/s at 1\,keV),
the reflection index, $R$, the ionization parameter, $\xi$, and a
factor to account for the relative normalization between PCU0 and
HEXTE cluster 0. If not stated otherwise, we plot 1$\sigma$ errors for
each parameter of interest. 

\vspace{-0.2cm}
\begin{table}[h]
\caption{Fixed parameters of the spectral model.}
\begin{center}
\begin{tabular}{lc}
\hline \hline
fix. parameter & value \\
\hline \hline
\textit{N}$_{\rm H}$ & 5$\cdot$10$^{22}$cm$^{-2}$\\
cutoff energy & no cutoff\\
redshift & 0\\
Z$>$2 element abundances & 1\\
iron abundance to abundance above & 1\\
inclination angle & 70\degs\\
power law index for reflection emissivity & -2\\
inner disk radius & 6\,GM/c$^2$\\
outer disk radius & 1000\,GM/c$^2$\\
\hline \hline
\end{tabular}
\label{fixedpara}
\end{center}
\end{table}
\vspace{-1.0cm}

\subsection{{\it GBI} and {\it RT}}

The appearance of steady radio emission in $\chi$-states suggested us
to search for correlations between {\it RXTE} data and radio data at
15\,GHz ({\it Ryle Telescope = RT}) and 2.25\,GHz ({\it Green Bank
Interferometer = GBI}). \grs\ shows variability at all frequencies on
time scales of seconds to hours. For a useful statement in
$\chi$-states a suitably selection of corresponding datasets is
therefore required.

According to the general interpretation that the radio emission is
synchrotron emission from ejected plasma in sporadic or continuous
jets (Fender \etal 1995) the radio flux should reach a maximum 15\,min
after the actual ejection (Mirabel \etal 1997) and therefore after a
possible determining X-ray event.

\begin{figure}[h]
\begin{center}
 \vbox{\psfig{figure=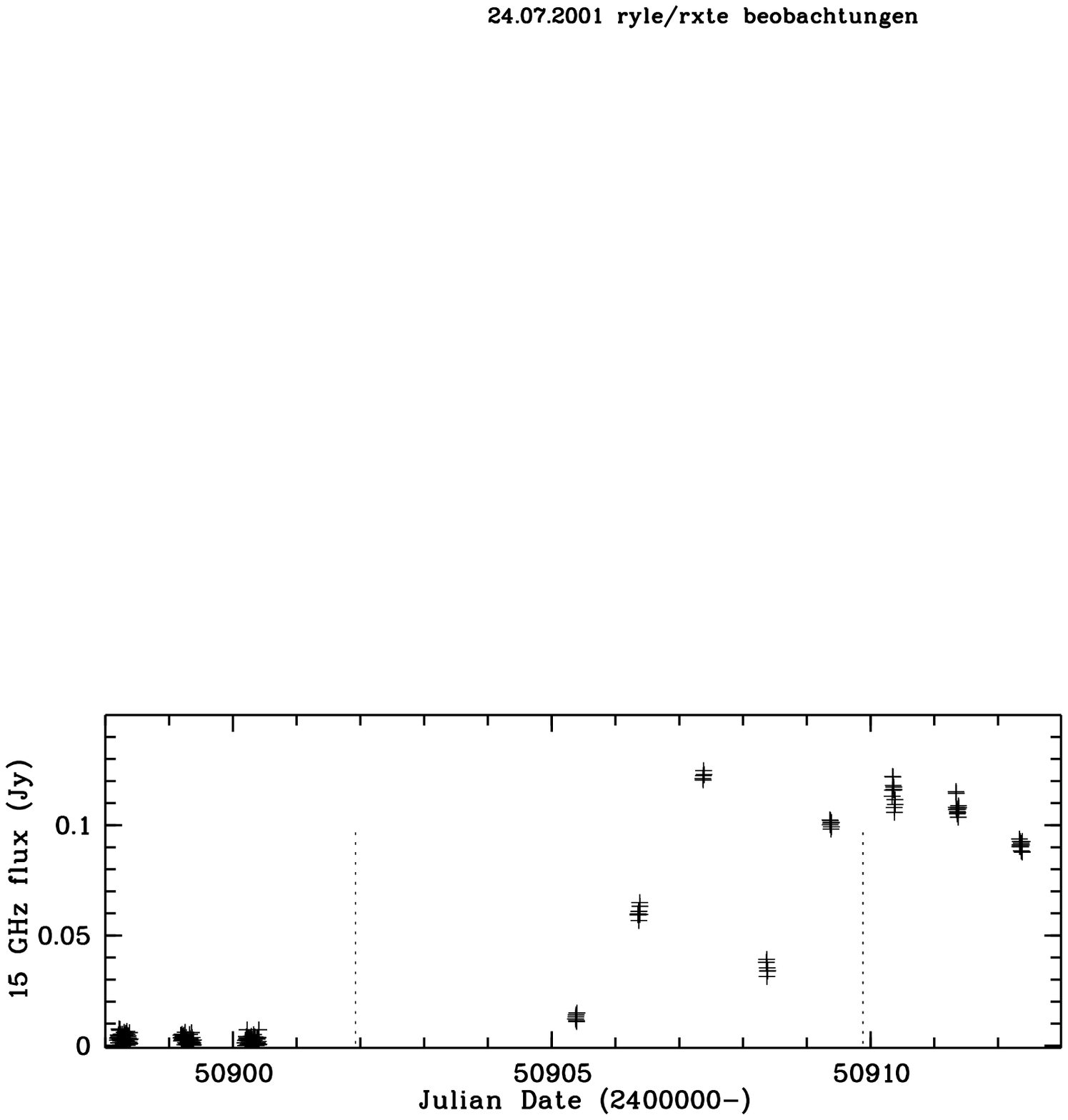,width=9.5cm,%
          bbllx=1.4cm,bblly=10.5cm,bburx=19.cm,bbury=17cm,clip=}}\par
 \caption[]{Variation of the 15\,GHz flux
          (crosses) between JD 2450898 and JD 2450913. The dotted
          lines mark the time of {\it RXTE} observations from 29.03.1998 (JD
          $\sim$2450902) and 06.04.1998 (JD $\sim$2450910). \grs\
          showed low radio emission until JD
          2450900.5. The source was dominated by strong, variable
          radio emission following JD 2450905. No statement can be made
          about the time in between, because of the lack of radio
          observations during the {\it RXTE} observation from 29.03.1998. Also
          for the {\it RXTE} observation from 06.04.1998 the radio flux is
          uncertain because of the strong variation before and after
          the X-ray exposure. 
\label{rylewahl}}
\end{center}
\vspace{-0.5cm}
\end{figure}

Fig.~\ref{rylewahl} shows the 15\,GHz radio flux, $F_R$, from the {\it
RT} for JD 2450898--2450913 together with two {\it RXTE}
observations. The {\it RT} observed \grs\ several times a day with
five minute exposures.

Because the radio exposure is much shorter than the X-ray exposure
($\sim$hour) it is important to select radio data simultaneous with
the {\it RXTE} observations. 37 of the 139 analyzed {\it RXTE}
observations have simultaneous {\it RT} data and 9 have simultaneous
{\it GBI} data.

\begin{figure}[h]
\begin{center}
 \vbox{\psfig{figure=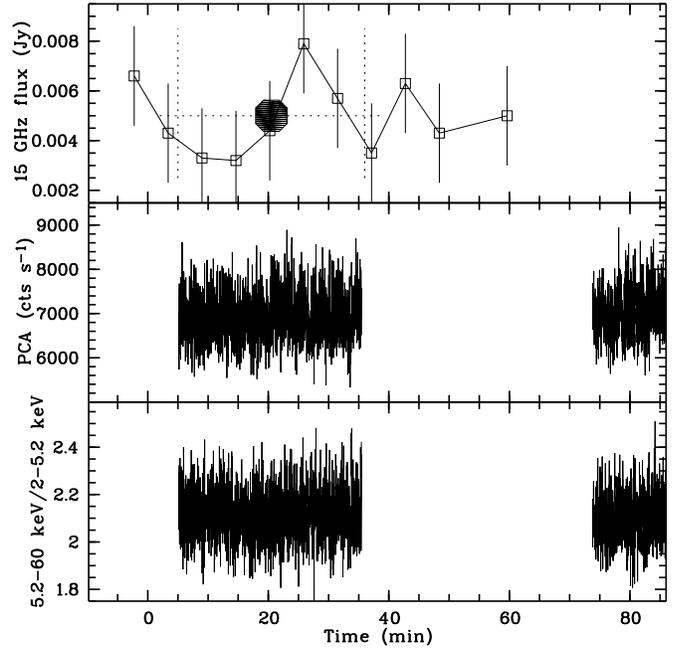,width=8.8cm,%
          bbllx=3.2cm,bblly=8.cm,bburx=16.6cm,bbury=21.8cm,clip=}}\par
 \caption[]{15\,GHz light curve ({\it RT}, upper pannel), 2--60\,keV PCA
          light curve (middle pannel) and hardness ratio
          $\frac{5.2-60 keV}{2-5.2 keV}$ (lower pannel) of the $\chi$-state from
          15.09.1998 (JD 2451071.9). While the radio light curve
          shows structured variability, none is seen in X-ray intensity
          and spectrum. The large black dot (marked for clarity with the dotted lines) in the upper panel represents the averaged radio flux over the RXTE observation used for further analysis. 
\label{ryleasm}}
% /local\_U/GRS1915/xte/work/reduction/rex/arbeit/lc/plot\_lc.prg
\end{center}
\vspace{-0.5cm}
\end{figure}

Still, the selection of simultaneous radio observations is
non-trivial. Occasionally the radio emission varies also during a
single $\chi$-state X-ray observation, whereas no variability is seen
in X-ray count rate and hardness ratio (5.2--60\,keV/2--5.2\,keV)
(Fig.~\ref{ryleasm}). But the variation of $F_R$ during a $\chi$-state
observation is negligible compared to the uncertainties of the
individual radio measurements and to the variation between different
{\it RXTE} observations. Therefore, the radio fluxes were averaged for
each {\it RXTE} observation.

\section{Results of the analysis}

\begin{figure}[h]
\begin{center}
 \vbox{\psfig{figure=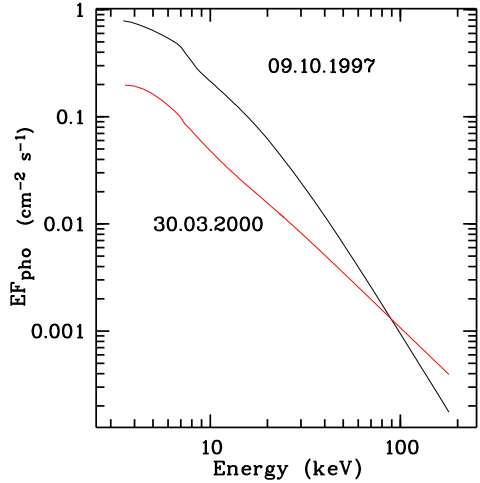,width=9.0cm,%
          bbllx=1.cm,bblly=10.cm,bburx=14cm,bbury=19.6cm,clip=}}\par
 \caption[]{Combined PCU0 and HEXTE cluster 0 spectra of the
          $\chi$-states from 09.10.1997 (down, artificially offset for better discrimination) and 30.03.2000
          (top). Theses spectra show extreme power law slope ($\Gamma$=3.41$\pm$0.06
          (09.10.1997), $\Gamma$=2.65$\pm$0.05 (30.03.2000) and reflection
          ($R$=7.43$\pm$1.12 (09.10.1997), $R$=0.35$\pm$0.23
          (30.03.2000) behavior. Note the obvious reflection hump for the
          09.10.1997 spectra at $\sim$10\,keV.    
\label{GR}}
\end{center}
\vspace{-0.9cm}
\end{figure}
% arbeit/example_spectra/plot_GR.prg

The spectra of all 139 datasets have been fitted with the
above-described DISKBB+REFSCH model. Two example model spectra are
shown in Fig.~\ref{GR}. These spectra have the extreme values of
$\Gamma$ and $R$, revealing the spectral differences between different
$\chi$-state observations.

\begin{figure*}[h]
\begin{center}
\psfig{figure=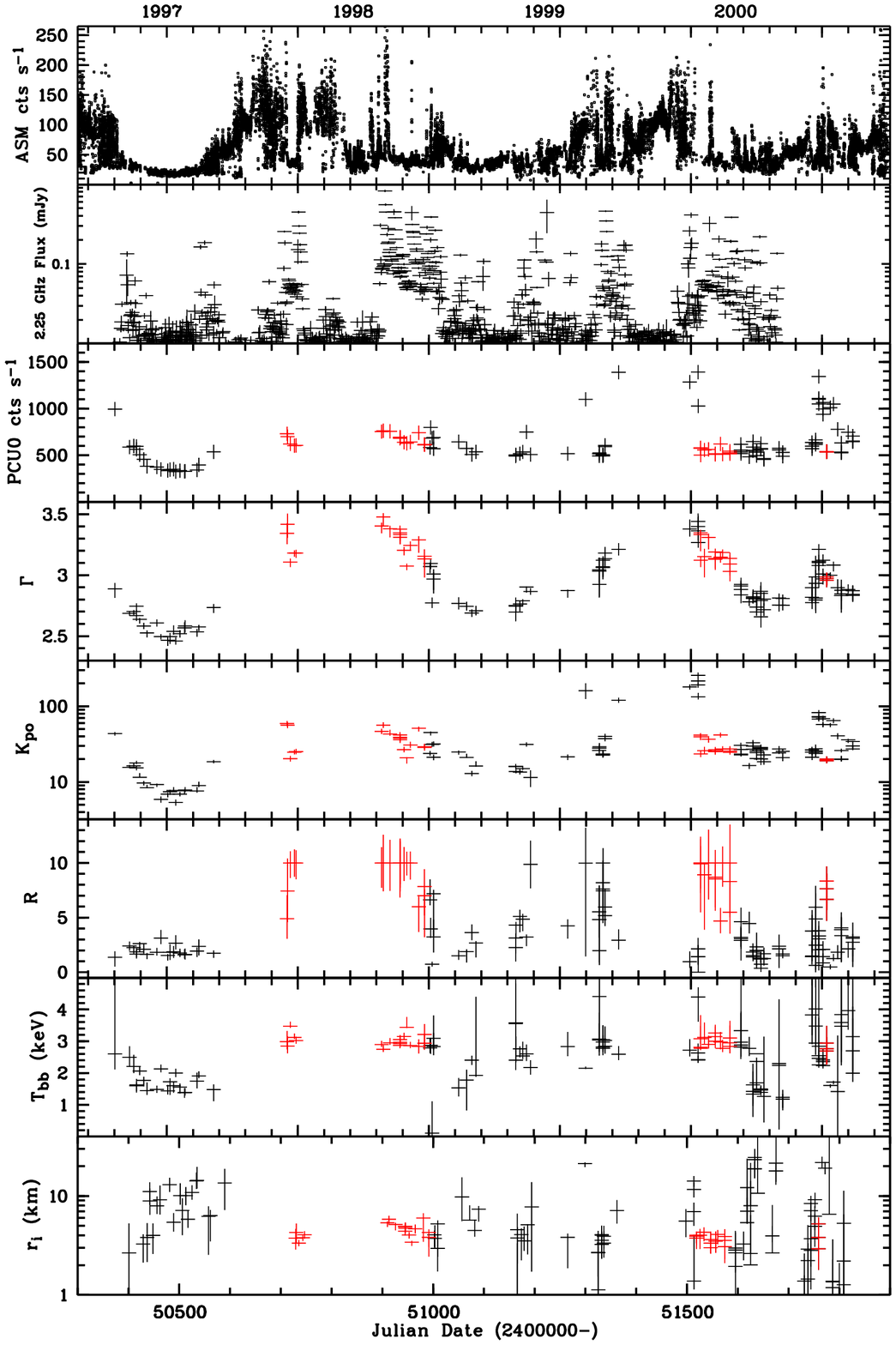,width=16.0cm,%
          bbllx=1.cm,bblly=3.cm,bburx=17.8cm,bbury=27.4cm,clip=}\par
 \caption[]{1.5--12\,keV ASM count rate (1st panel from top) and 2.25\,GHz {\it GBI}
          flux (2nd) from November 1996 to September 2000. ({\it GBI}
          was off-line some time before JD $\sim$2450400 and following
          JD $\sim$2451660. The lower panels show the 3--20\,keV PCU0
          count rate (3rd), the power law slope, $\Gamma$, (4th) and
          power law normalization, $K_{po}$, (5th), the reflection
          amplitude, $R$, (6th), the accretion disk temperature,
          $T_{bb}$, (7th) and the inner disk radius (determined from
          the disk normalization), $r_i$, (8th) for all analysed
          $\chi$-states of \grs. Error bars are 1$\sigma$ for one parameter of interest. The observations of the two different branches in the $\Gamma$($K_{po}$) behavior are marked with red (steep) and black (flat) as in Fig.~\ref{alpK}.   
\label{fitres1}}
\end{center}
\end{figure*}

The best fit parameters for the $\chi$-states of \grs\ are shown in
Fig.~\ref{fitres1} and Table~\ref{fitrestab}, respectively. The upper
two panels in Fig.~\ref{fitres1} show the strong variability of the
source in the 1.5-12\,keV ({\it RXTE} All Sky Monitor (ASM)) and
2.25\,GHz ({\it GBI}) bands from JD 2450300--2451900. Irregular
outburst and relatively quiet phases alternate in X-ray and radio
without any obvious coupling.

The lower six panels of Fig.~\ref{fitres1} show the PCU0 count rate
and the fit parameters of the $\chi$-state observations. The red and
black data points mark two groups of observations with different
$\Gamma$($K_{po}$) behavior (see Fig.~\ref{alpK}).

Between the $\chi$-states, the PCU0 count rate varies, with some
exceptions, by a factor of 2 at most. Only during the long continuous
$\chi$-state at JD $\sim$2450400--2450600 and at JD $\sim$2451750
recurrent variability in the X-ray count rate is observed. The model
parameters of DISKBB+REFSCH are variable. The power law slope,
$\Gamma$, varies between 2.4 and 3.5 with a long-term periodicity of
$\sim$590\,days (Rau \& Greiner 2002, in preparation). No correlation
of $\Gamma$ and PCU0 count rate is seen, except around JD
$\sim$2450500 and JD $\sim$2451750. The power law normalization
behaves similarly to the slope.

The reflection component, $R$, is variable between different
$\chi$-state observations and shows a long-term variability similar to
the power law slope. It varies between 0 and 10 (upper limit of our model) with rather large
uncertainties when $R>$4. Except for five observations when $R<$1,
this rules out an isotropically sandwiching corona above the entire accretion disk. 

The significance of the accretion disk component varies through the
$\chi$-states. For some observations the disk is more or less absent
(e. g. JD 50400--50600; relatively low disk temperature and large
inner disk radius (small normalization)), for other observations the
disk component provides a non-negligible contribution to the X-ray
flux (e. g. JD 51500--51600; high temperature and small inner
radius). In order to be consistent and to fit all $\chi$-state
observations with the same model, we included the DISKBB component in
all of our fits, although it could be excluded for several observations.

\begin{table*}[t]
\caption{Fit results of the {\it RXTE} spectra using the DISKBB+REFSCH
model (The complete table is available at CDS). (1): ID of observation
(I=10402-01, J=20187-02, K=20402-01, L=30182-01, M=30402-01,
N=30703-01, O=40703-01, P=50703-01), (2): exposure time of
observation, (3): 3--20\,keV PCU0 count rate, (4): 20--190\,keV HEXTE
cluster 0 count rate, (5): accretion disk temperature, (6): accretion
disk normalization ($K_{bb}=(\frac{r_i}{D/10~kpc})^2\cos\theta$), (7):
power law slope, (8): reflection amplitude, (9): power law
normalization (photons/keV/cm$^2$/s at 1 keV), (10): ionization
parameter, (11): factor for normalization of PCU0 and HEXTE cluster 0,
(12): reduced $\chi^2$. All errors are 1$\sigma$ for each parameter of
interest.}
\begin{tabular}{lccccccccc}
\hline \hline
Obs-ID$^{(1)}$ & GD & JD & Exposure$^{(2)}$ & PCU0$^{(3)}$ & HEXTE0$^{(4)}$ & T$_{bb}$$^{(5)}$ & K$_{bb}$$^{(6)}$ & $\Gamma$$^{(7)}$ & R$^{(8)}$ \\
& & (-2400000) & [s] & [cts/s] & [cts/s] & [keV] & & &\\
\hline \hline
K-02-01 & 14.11.1996  & 50401.12 & 2688 & 995 & 70 & 2.60$^{+4.81}_{-0.49}$  & 1.67$^{+4.90}_{-1.60}$ & 2.89$^{+0.05}_{-0.07}$  & 1.37$^{+0.55}_{-0.86}$\\ 
K-06-00 & 11.12.1996  & 50428.86 & 8656 & 587 & 60 & 2.50$^{+0.35}_{-0.30}$  & 2.52$^{+1.46}_{-1.45}$ & 2.69$^{+0.02}_{-0.02}$  & 2.40$^{+0.40}_{-0.44}$\\ 
K-07-00 & 19.12.1996  & 50436.74 & 8992 & 598 & 61 & 2.21$^{+0.36}_{-0.21}$  & 3.42$^{+3.25}_{-2.32}$ & 2.70$^{+0.01}_{-0.03}$  & 2.25$^{+0.20}_{-0.56}$\\ 
. & .&. & .&. & .&.&.&.&. \\ 
. & .&. & .&. & .&.&.&.&. \\ 
P-28-00 & 21.09.2000  & 51808.70 & 2384 & 706 & 69 & 2.70$^{+1.39}_{-0.73}$  & 1.14$^{+8.85}_{-1.01}$ & 2.84$^{+0.03}_{-0.06}$  & 3.11$^{+1.42}_{-2.64}$\\ 
P-28-01 & 21.09.2000  & 51808.77 & 2400 & 651 & 66 & 3.15$^{+3.59}_{-1.05}$  & 0.38$^{+0.21}_{-0.37}$ & 2.83$^{+0.08}_{-0.04}$  & 2.74$^{+0.22}_{-1.29}$\\ 
P-28-02 & 21.09.2000  & 51808.83 & 2336 & 645 & 44 & 2.00$^{+1.72}_{-0.28}$  & 6.65$^{+23.43}_{-6.47}$ & 2.87$^{+0.05}_{-0.03}$  & 3.24$^{+0.24}_{-2.29}$\\ 
\hline \hline
\end{tabular}
\label{fitrestab}
\end{table*}

\begin{table}[h]
\begin{tabular}{lcccc}
\hline \hline
Obs-ID  & K$_{po}$$^{(9)}$ & $\xi$$^{(10)}$ & c$^{(11)}$ & $\chi^2$$^{(12)}$ \\
 & [Pho/keV/cm$^{2}$/s] & & & \\
\hline \hline
K-02-01 & 43.5$^{+2.0}_{-2.6}$  & 4999$^{+1}_{-3521}$ & 0.82 & 0.69\\ 
K-06-00 & 15.6$^{+1.2}_{-0.3}$  & 5000$^{+0}_{-4421}$  & 0.78 & 1.00\\ 
K-07-00 & 16.3$^{+1.1}_{-0.4}$  & 996$^{+4004}_{-338}$  & 0.79 & 1.01\\ 
.&. & .&.&. \\ 
.&. & .&.&. \\ 
P-28-00 & 34.1$^{+0.7}_{-5.0}$  & 2902$^{+2098}_{-2902}$  & 0.82 & 0.94\\ 
P-28-01 & 29.8$^{+0.1}_{-2.1}$  & 462$^{+4538}_{-461}$  & 0.81 & 0.84\\ 
P-28-02 & 27.2$^{+2.8}_{-0.8}$  & 34$^{+2900}_{-34}$  & 0.52 & 0.77\\ 
\hline \hline
\end{tabular}
\end{table}

The accretion disk component shows a variable disk temperature of
1--4\,keV. Sometimes, large uncertainties due to the small
contribution of the DISKBB component to the total flux are seen. The
inner disk radius, which can be determined from the disk
normalization, varies between 1 and 20\,km. For a non-rotating black
hole of mass 14\,\msun\ (as measured for \grs, Greiner \etal 2001b)
the Schwarzschild radius is $\sim$40\,km. It is known that the DISKBB
model underestimates the inner disk radius by a factor of 1.7--3 due
to Doppler blurring and gravitational redshift (Merloni \etal
2000). Also the neglect of comptonization in the surface layers of the
disk leads to unphysical values when using then DISKBB model
(Zdziarski \etal 2001). But even a maximally rotating black hole
(where the inner disk radius reaches the Schwarzschild radius) cannot
account for the majority of the parameter values. This problem has to
be kept in mind when discussing the absolute values of the parameters.

Another free physical parameter of the REFSCH model is the ionization
parameter, $\xi$. It has huge uncertainties because no Fe K$\alpha$
line could be fitted but has a negligible influence on the hard X-ray
continuum and our model parameters. Therefore we will not plot or
discuss $\xi$ further. Theoretically, the reflection in the hard
spectrum should depend on the ionization parameter because higher
ionization means lower absorption and therefore higher reflection
probability. But the REFSCH model includes a simple 1-ionization zone
model only, which is very unlikely to be present in the disk and does
not show any dependence of reflection and ionization at all.

\subsection{The power law component}

\begin{figure}[h]
\begin{center}
 \vbox{\psfig{figure=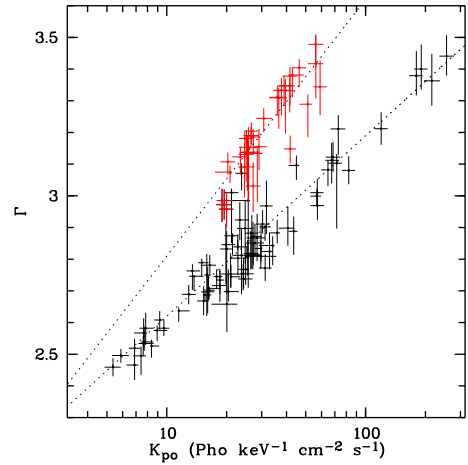,width=9.5cm,%
          bbllx=1.5cm,bblly=9.5cm,bburx=19.cm,bbury=21.5cm,clip=}}\par
 \caption[]{Power law slope, $\Gamma$, as a function of power law
          normalization, $K_{po}$. Each {\it RXTE} observation is
          represented by one data point. The upper branch presents the red points in Fig.~\ref{fitres1} and the lower
          branch the black points, respectively. The dotted lines
          represent the fitted correlation functions (see text).
\label{alpK}}
\end{center}
\vspace{-0.5cm}
\end{figure}

The model fits reveal an increasing power law normalization, $K_{po}$, with a steepening power law component (Fig.~\ref{alpK}) suggesting a pivoting behavior. Two branches with different slopes are seen in the correlation. No correlation of the power law slope with the X-ray count rate in ASM and/or PCU is observed. 

The strength of the correlation can be tested using a Spearman
rank-order correlation test (Press \etal 1992). Both branches show
strong correlations (steeper: $r_{S}$=0.85, flatter:
$r_{S}$=0.87). (Note, this statistical method does not take into
account the particular uncertainties of the data points.)

The best descriptions of the correlations are
functions of the type 

\begin{equation}
\Gamma=u\cdot logK + v 
\end{equation}

with \textit{u}=0.57$\pm$0.02 and \textit{v}=2.0$\pm$0.1 for the lower branch and \textit{u}=0.81$\pm$0.07 and \textit{v}=2.0$\pm$0.1 for the upper branch.

\begin{figure}[h]
\begin{center}
 \vbox{\psfig{figure=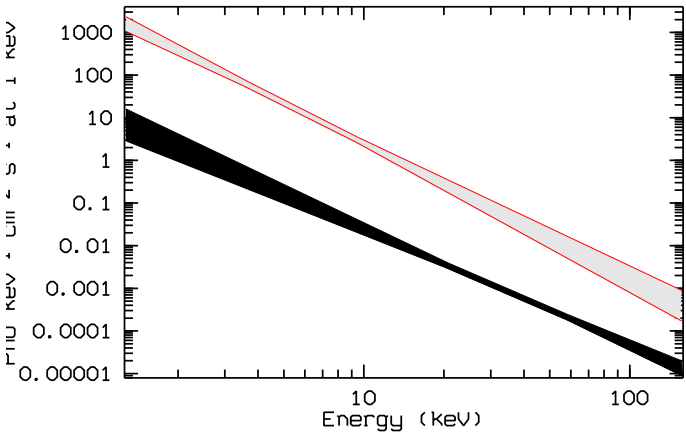,width=9.5cm,%
          bbllx=3.cm,bblly=2.cm,bburx=16.5cm,bbury=10.5cm,clip=}}\par
 \caption[]{Schematic power law spectra for the upper (top) and lower (bottom) branch
          of Fig.\ref{alpK}, the red and black points from
          Fig.~\ref{fitres1}, respectively, artificially offset by a
          factor of 100. The pivoting happens at
          different energies (4--8\,keV \& 20--30\,keV). 
\label{pivote}}
\end{center}
\vspace{-1.0cm}
\end{figure}

The pivoting has been tested in more detail by schematically plotting the obtained power law spectra. The two different branches consequently show different pivoting
behavior (Fig.~\ref{pivote}). Whereas the upper branch from
Fig.~\ref{alpK} shows a pivoting energy of 4--8\,keV (gray), the lower branch
pivots at around 20--30\,keV (black).

\subsection{Reflection}

Due to the known uncertainties of the PCA below 6\,keV (Jahoda \etal
1996) and the low energy resolution, stringent conclusions about the
behavior of an iron line in the {\it RXTE} spectra at 6.4--7\,keV not
can be drawn. For the analysis of the reflection, therefore, only the
continuum radiation can be used, manifesting itself in the reflection
hump from 10--30\,keV.

\begin{figure}[h]
\begin{center}
 \vbox{\psfig{figure=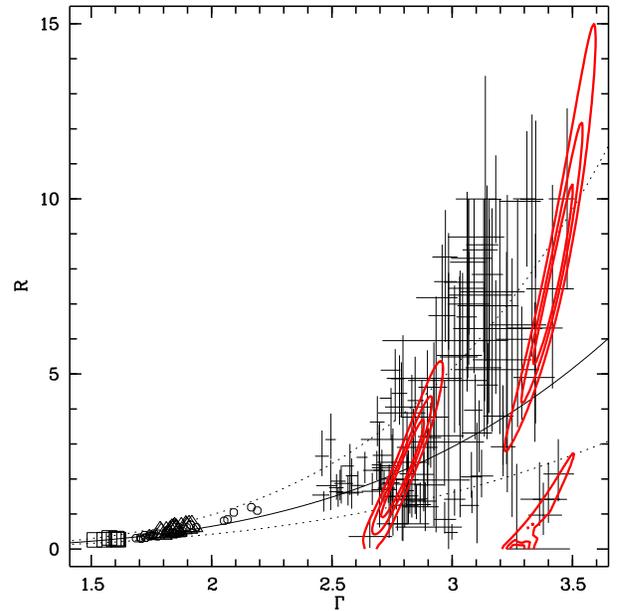,width=9.5cm,%
          bbllx=1.cm,bblly=1.8cm,bburx=16.5cm,bbury=15.5cm,clip=}}\par
 \caption[]{$R$($\Gamma$)-correlation for \grs\ (crosses) and three X-ray binaries (Cyg X-1 = circles, GX 339-4 = triangles and GS 1354-644 =
          squares; from Gilfanov \etal 2000). The solid line
          represents the best fitting model function and the dotted
          lines the upper and lower limits. Overplotted are the 1, 2
          and 3\,$\sigma$ confidence contours for three different
          observations (from left: N-23-01, K-49-01, O-40-03). Note the data points (JD 2451497--2451513) at $\Gamma\sim$3.3--3.5 and $R\sim$0--2
          behaving remarkably different.
\label{bhc-corr}}
\end{center}
\end{figure}

The reflection amplitude in \grs\ as a function of $\Gamma$ is shown
in Fig.~\ref{bhc-corr} together with that of three black hole
candidates (Cyg X-1, GX 339-4 and GS 1354-644; Gilfanov \etal
2000). Although $R$ shows large uncertainties for the $\chi$-states of
\grs, a similar correlation is seen as in the other X-ray
binaries. The steeper the power law component, the higher is the
reflected fraction of photons. Fig.~\ref{bhc-corr} contains the
confidence intervalls of $R$ for three observations to clarify the
influence of the model on the $R$($\Gamma$)-correlation. Although the
contours are elongated similiar to the correlation (higher $\Gamma$
has higher $R$), they are intrinsically steeper compared to the
overall correlation. Thus, the existence of the
$R$($\Gamma$)-correlation in the $\chi$-states of \grs\ is no artifact
of the model.

A group of observations at high $\Gamma$ (3.3--3.5) and small $R$
(0--2) behaves remarkably differently. These points belong to five
datasets from JD 2451497 and JD 2451513 showing very high PCU0 count
rates (1300--2100 cts/s) and are short duration $\chi$-states between
different high variability states.

The correlation of $R$ and $\Gamma$ is tested using a Spearman
rank-order correlation test. The correlation is distinct
($r_{s}$=0.61) but not strong.

For a quantitative description of the correlation, a phenomenological
function was fitted to the data. The best-fitting model is a power law

\begin{equation}
R= u \cdot \Gamma^v,
\end{equation}

with \textit{u}=0.05$\pm$0.007 and \textit{v}=3.7$\pm$0.4 (solid line
in Fig.~\ref{bhc-corr}).

\subsection{X-ray-Radio-Correlation}

No obvious correlation between the soft X-ray component and the radio
emission is seen. Neither the soft X-ray flux, the temperature of the
accretion disk nor the inner disk radius show a correlation with the
15\,GHz radio flux from {\it RT} and/or {\it GBI} .

\begin{figure}[h]
\begin{center}
 \vbox{\psfig{figure=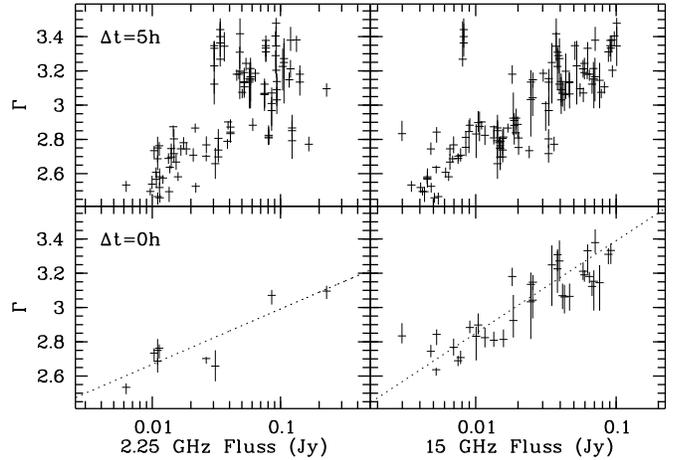,width=9.5cm,%
          bbllx=1.3cm,bblly=14.cm,bburx=19cm,bbury=25.5cm,clip=}}\par
 \caption[]{$\Gamma$($F_R$)-correlation for {\it GBI} (left) and {\it RT}
          (right). Upper panels: correlation for $\Delta t$=5\,hrs
          (time between {\it RXTE} and radio observation). As for {\it GBI} and
          for {\it RT} the correlation is visible but the data points show a large spread. Lower panel: correlation for
          $\Delta t$=0\,hrs. The dotted lines represent the correlation
          functions. 
\label{alpharadio}}
\end{center}
\vspace{-0.8cm}
\end{figure}

An unexpected result is found when plotting $\Gamma$ vs. $F_R$
(Fig.~\ref{alpharadio}). The power law slope correlates positively
with the radio flux at 2.25\,GHz and 15\,GHz. Observations with high
radio emission show a softer X-ray spectrum (steeper power law
component). Note that no bimodality in the radio emission
exists. Instead, a continuous spread is observed. A strict separation
of radio loud and radio quiet $\chi$-states, as done before
(e. g. Muno \etal 2001, Trudolyubov 2001) seems therefore
unsubstantiated.

The correlation of $\Gamma$ and F$_R$ is most obvious for simultaneous
($\Delta t$=0\,hrs) observations. Datasets with radio observations
$\pm$5\,hrs offset also show the correlation (Fig.~\ref{alpharadio}
upper panel) but with significant scattering. Usually, no statement
about the variability state before and after the {\it RXTE} exposure
can be made, because the source may have had several state
alterations. Thus, non-simultaneity is the likely reason for the
strong scattering of the correlation for $\Delta t$=5\,hrs in
comparison to $\Delta t$=0\,hrs.

To determine the strength of the correlation a Spearman rank-order
correlation test was made for the {\it RT} data with $\Delta
t$=0\,h. The correlation is strong ($r_s$ = 0.83).  No test was made
for {\it GBI} due to the small number of simultaneous {\it GBI}/{\it
RXTE} datasets.
 
The next step was to fit a phenomenological function (dotted line in
Fig.~\ref{alpharadio}) to the correlation. The best model functions
are of the form

\begin{equation}
\Gamma=u\cdot log F_R + v,
\end{equation}

with \textit{u}=0.54$\pm$0.02 and \textit{v}=3.94$\pm$0.03 for {\it
RT} and \textit{u}=0.32$\pm$0.05 and \textit{v}=3.3$\pm$0.1 using {\it
GBI} data.

\begin{figure}[h]
\begin{center}
 \vbox{\psfig{figure=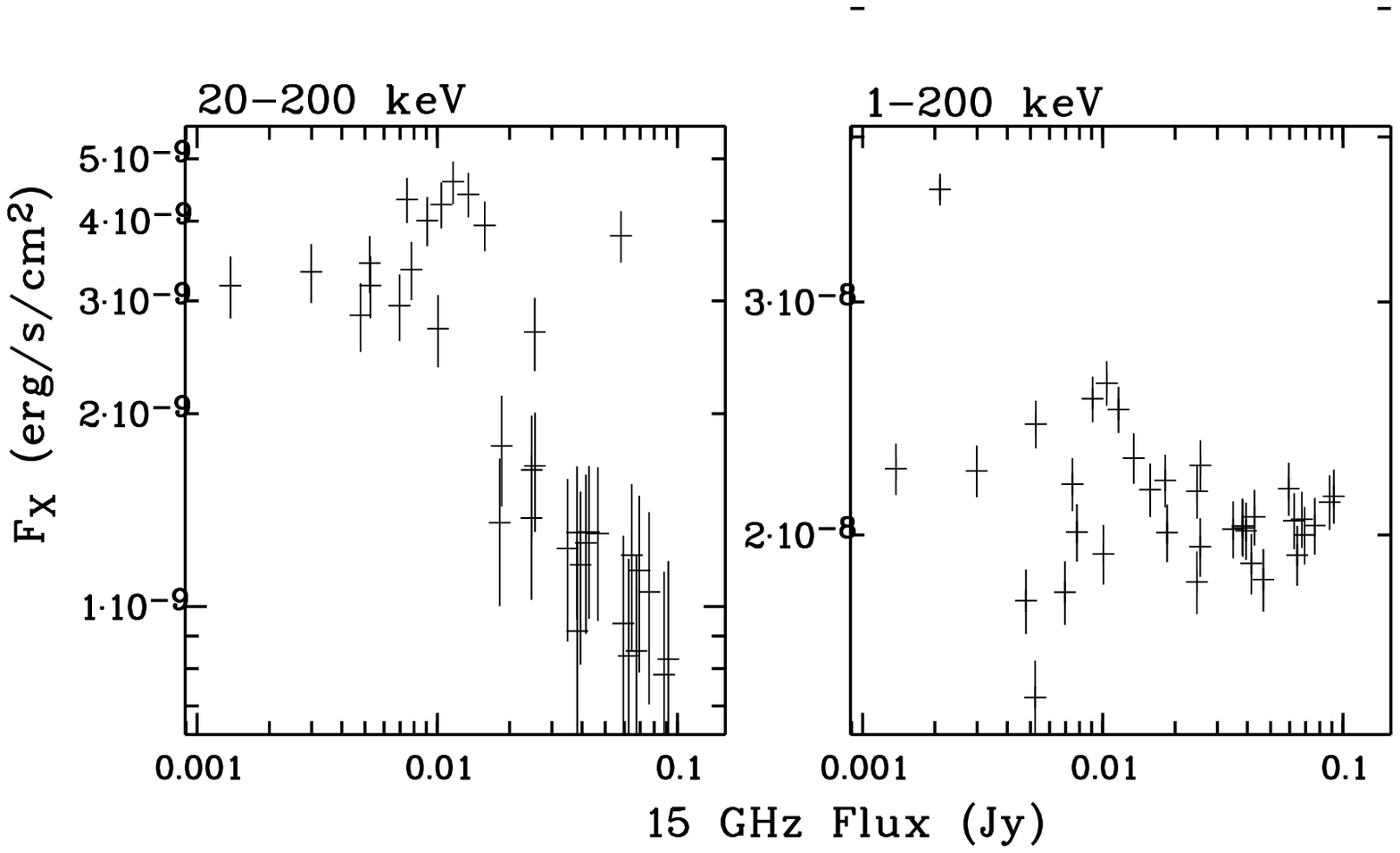,width=9.5cm,%
          bbllx=1.5cm,bblly=18.cm,bburx=19.5cm,bbury=27cm,clip=}}\par
 \caption[]{20--200\,keV (left) and 1--200\,keV (right) X-ray flux
          shown over the 15\,GHz radio flux ({\it RT}). The 20--200\,keV
          flux has a negative correlation with the radio flux, whereas
          the 1--200\,keV flux does not show such a correlation. 
\label{xryleflux}}
\end{center}
\vspace{-0.8cm}
\end{figure}

Besides the $F_R$($\Gamma$)-correlation we also investigated the
relation between $F_R$ and the X-ray flux of the power law
component. This introduces some freedom as to at which lower energy
the power law is ``chosen''. Fig.~\ref{xryleflux} shows the X-ray flux
from 20--200\,keV and 1--200\,keV, respectively, over the radio flux.

With increasing radio emission, the X-ray flux in the power law
component between 20--200\,keV decreases (neglecting reflection). The
total 1-200\,keV X-ray flux in the power law component is more or less
constant. No correlation with the radio flux is seen. This is because
the flux in the 1--20\,keV range by far dominates, thus washing out
the correlation. Muno \etal (1999) did not find a correlation of
$F_X$(50--100\,keV) with the 15\,GHz radio flux. But they used a model
consisting only of DISKBB and a broken power law (BKNPO) for the X-ray
spectra and {\it RT} data with$\pm$12\,hrs offset.

A Spearman rank-order correlation test for $F_X$(20--200) gives
$r_s$=-0.75, thus showing a strong anti-correlation. For
$F_X$(1--200\,keV) the test results in $r_s$=-0.09 and ratifies the
lack of a correlation between the 1--200\,keV non-thermal X-ray flux
and $F_R$.

\begin{table}[h]
\caption[]{Observed correlation of the $\chi$-states of \grs.
\label{korrelationen}}
\begin{center}
\begin{tabular}{cccc}
\hline \hline
\multicolumn{2}{c}{correlation between:} & type$^{(1)}$ & ref.\\
\hline \hline
$\Gamma$ & $K_{po}$ & + & \\
$\Gamma$ & $F_R$ & + &  \\
$\Gamma$ & $R$ & + & \\
$F_X$(20--200\,keV) & $F_R$ & - & \\
$\nu_{QPO}$(0.5--10\,Hz)$^{(2)}$  & $F_R$ & - & Muno \etal 2001 \\
$\nu_{QPO}$(0.5--10\,Hz)  & $T_{bb}$ & + & Muno \etal 1999 \\ 
\hline \hline
\end{tabular}
\end{center}
(1):
direction of correlation ("+" = positive, "-" = negative), (2):
frequency of the 0.5--10\,Hz QPO.
\vspace{-0.8cm}
\end{table}

\section{Discussion}

Table~\ref{korrelationen} summarizes the correlations found here plus
2 relevant correlations found by Muno \etal (1999, 2000). In the
following we will discuss implications of these correlations.

\subsection{Comptonization}

The origin of the hard power law component in AGN and X-ray binaries
is still under discussion. Three main models exist, where the hard
photons originate by inverse Compton scattering of soft disk photons
on hot electrons. The main difference between these models is the
distribution of the electrons. They can be thermal (Maxwellian),
non-thermal (power law like) or free falling from the last stable
orbit onto the event horizon of the black hole. The high energy
spectrum presents an important test for the distinction between these
models. Therefore, the behavior of the power law component in \grs\ is
crucial for the understanding of the electron distribution and for the
geometry of the system.

It is still unclear whether or not the hard spectrum of \grs\ in the
$\chi$-state extends with or without cutoff up to MeV energies. Due to
the rapid variability of the source and the long required exposure
times it is nearly impossible to get sufficiently accurate
$\gamma$-spectra with the present high energy satellites. For example,
the OSSE spectrum from 14th--20th of May 1997 reveals a power law tail
up to 1\,MeV without a cutoff (Iyudin 2000, Zdziarski \etal 2001). But
as several contemporaneous {\it RXTE} observations show (which cover
$\lax$3\% of the OSSE exposure), during the observation \grs\ went
through several $\chi$/$\alpha$ state transitions and the PCA photon
count rate from 2--40\,keV varied between 5 and 25 kcts/s. Thus, the
OSSE spectrum is the sum of the spectra of the $\chi$ and $\alpha$
(and possible other) states, with an unkown fraction coming from the
$\chi$-state. It is therefore premature to unequivocally relate the
hard MeV tail to the $\chi$-state.

\subsubsection{Bulk Motion Comptonization}

\begin{figure}[h]
\begin{center}
 \vbox{\psfig{figure=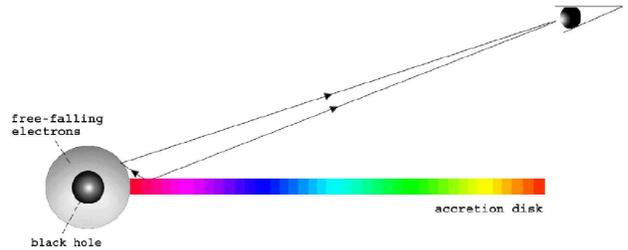,width=9.cm,%
          bbllx=1.cm,bblly=1.cm,bburx=20.5cm,bbury=10.5cm,clip=}}\par
 \caption[]{Bulk motion geometry. Electrons become spherical
          free-falling between the last stable orbit and the event
          horizon due to shocking. The photons are comptonized on
          these accelerated electrons.
\label{bulkmotion}}
\end{center}
\vspace{-0.8cm}
\end{figure}

One model for the comptonization is the bulk motion comptonization
(Blandford \& Payne 1981, Chakrabarti \& Titarchuk 1995)
(Fig.~\ref{bulkmotion}). The accretion stream passes the transition
radius, $r_{ts}$, at the last stable orbit around the black hole and
gets shocked. The matter falls spherical onto the event horizon. In
front of the shock, the stream is similar to an optically thick,
geometrically thin accretion disk (as required by the strong observed
blackbody component).

The free falling electrons are accelerated up to the speed of light
and inverse Compton scattering of soft photons on the electrons
provides the observed power law in the hard spectrum. The model
predicts a cutoff depending on the mass accretion rate (Ebisawa \etal
1996) due to inverse Compton scattering and Compton recoil. As stated
above, the existence/absence of a cutoff during the $\chi$-states is
still not secured, wherefore not statement about the model can be made
from that issue.

Two major problems exist for the description of $\chi$-state data by
the BMC model. The {\it RXTE} spectra of \grs\ partly show a strong
reflection component. This is not compatible with the geometry of a
thin disk outside a central, spherical accretion stream. The accretion
stream is accelerated away from the inner disk edge onto the event
horizon of the black hole. Therefore only a small part of the soft
photons are comptonized back into the disk plane at $r>r_{ts}$. In
conflict to the observations, no strong reflection component can
emerge. A possible explanation for the observed reflection can then be
a partly covering of the disk by an absorbing medium with
$\tau_T\sim$3 (Zdziarski 2000). Recent Chandra X-ray spectra indeed
show a high X-ray column density and abundance excesses for Si and Fe,
which may be related to material that is associated with the immediate
environment of the source (Lee \etal 2002).

The other problem for the BMC model is the time lag of hard and soft
X-ray photons. Muno \etal (2001) found a phase lag in the 5\,Hz QPO
(which are connected to the hard X-ray photons) of 0.5. This
corresponds to a delay of 0.1\,s of the hard photons to the soft
photons. The expected delay for scattering on a convergent electron
stream inside $r_{ts}$ is $\sim$0.2\,ms, much smaller than
observed. The phase lag is not constant in time and frequency and
seems to vary with the radio flux.

It is obvious that the BMC model not can explain the hard power law
component in the $\chi$-states of \grs. There are indications that the
BMC model fits other variability states. Shrader and Titarchuk (1998)
found good spectral agreement of the BMC model and {\it RXTE}
observations in the $\rho$-state. Note, however, that they completely
neglected the time delay of hard and soft photons and the strong
variability of the $\rho$-state for their X-ray spectral modelling.

\subsubsection{Thermal Corona}

\begin{figure}[h]
\begin{center}
 \vbox{\psfig{figure=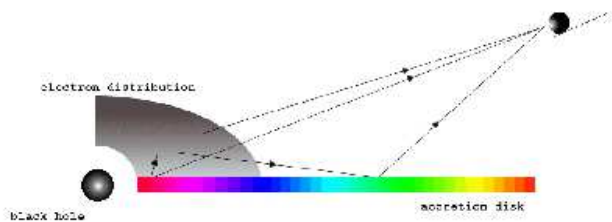,width=9.cm,%
          bbllx=3.5cm,bblly=12.0cm,bburx=17.8cm,bbury=17.8cm,clip=}}\par
 \caption[]{Disk corona geometry. The disk photons are comptonized in
          an electron distribution located above the accretion
          disk. (Note, the shape and size of the electron distribution in this plot has no real physical or observational meaning.) Part of the photons are scattered back into the disk plane
          and are reflected or reprocessed. The photons
          reaching the detector are compound of the soft
          disk photons, the comptonized and the reflected photons. 
\label{corona}}
\end{center}
\vspace{-0.8cm}
\end{figure}

Another possible model for the origin of the hard X-ray component is
the disk corona geometry (Fig.~\ref{corona}) (e.g. Haardt \& Maraschi
1993, Svensson \& Zdziarski 1994). Soft disk photons are inverse
Compton scattered on hot ($>$20\,keV) electrons located above the
disk.

Simultaneous {\it CGRO}/OSSE and X-ray observations of X-ray binaries
with black holes showed that the high energy continuum in the low/hard
state originates most likely due to thermal comptonization. The
predicted cutoff in the hard X-ray spectrum is seen in \cyg\
(Gierli\'nski \etal 1997), in the LMXB \gx\ (Zdziarski \etal 1998) and
some other X-ray transients (Grove \etal 1998).

Also the high/soft state in X-ray binaries can be explained by thermal
comptonization but requires high electron temperatures in some systems
to explain the observed unbroken power law up to MeV. In addition
$\tau_T\ll$1 is required in order to keep the spectrum soft (Zdziarski
2000).

The power law slope is a measure of the Compton amplification,
$A(L_{soft}+L_{diss})/L_{soft}$. $L_{diss}$ is the dissipated energy
in the corona and $L_{soft}$ the energy of the incoming
photons. Table~\ref{gamma-A} shows how $A$ depends on $\Gamma$ with

\begin{equation}
\Gamma=\frac{7}{3}(A-1)^{-\delta}
\end{equation}

\noindent (Beloborodov 1999) and $\delta$=1/6 for galactic black
holes. The ratio of the dissipated energy in the corona and the energy
of the intercepted soft photons gets smaller with a steeper
spectrum. For the highest $\Gamma$ only $\sim$9\% of the soft photon
energy is dissipated into the corona.

\begin{table}[h]
\caption[]{Fraction of the dissipated energy in the corona,
$L_{diss}$, on the energy of the incoming photons, Compton parameter,
$y=4\tau_T^2kT_e/(m_ec^2)$ ($\tau_T$ is the Thomson depth), and Compton amplification, $A$, for three different power law slopes.
\label{gamma-A}}
\begin{center}
\begin{tabular*}{8cm}{@{\extracolsep{\fill}}cccc}
\hline \hline
$\Gamma$ & $L_{diss}$ & $y$ & $A$ \\
\hline \hline
2.4 & 84\,\% & 1.33 & 1.84 \\
3.0 & 22\,\% & 3.65 & 1.22 \\
3.5 & 9\,\% & 7.30 & 1.09 \\
\hline \hline
\end{tabular*}
\end{center}
\end{table}

The minimum energy which a photon receives when passing a thermal
electron distribution depends on the electron temperature $kT_e$

\begin{equation}
\frac{\Delta \epsilon}{\epsilon}= \frac{4kT_e}{m_ec^2}.
\end{equation}

The result is a lack of low energy photons when $kT_e$ becomes
large. For $kT_e$$>$200\,keV the discrete orders (mainly the first) of
the Compton scattering become visible as an additional peak overlaying
the soft disk component. The high disk temperatures (3--4\,keV) in
\grs\ provide a photon deficit below 6--8\,keV. This makes obvious, as
already mentioned above, that a simple power law model is insufficent
for the description of the comptonized photons. The lack of photons at
low energies affects $\Gamma$, $K_{po}$ and the reflection component
in the model.

\subsubsection{Hybrid Corona}

The radiation processes and geometry can also be described by
comptonization of soft disk photons on a hybrid (thermal and
non-thermal) electron distribution above an optical thick accretion
disk (as in Fig.~\ref{corona})(Coppi 1992). Part of the electrons in
the thermal corona are accelerated, possibly in reconnection
events. The disk photons are comptonized on these electrons forming
the observed power law component in the hard X-ray spectrum. The
energy of the non-thermal electrons is partially transfered to the
thermal electrons due to Coulomb scattering. This leads to heating of
the thermal electrons above the Compton temperature. Then the thermal
electrons play an important part in the up-scattering of the soft disk
photons, too.

The compactness (ratio of luminosity and size) is an important
parameter of the coronal plasma (Coppi 1999). A large compactness
leads to electron positron pair creation due to photon photon
collisions resulting in a pair annihilation line at 511\,keV in the
spectra. When the compactness is low, the loss of energy of the
electrons due to Coulomb scattering dominates the loss due to inverse
Compton scattering. The ratio of thermal to non-thermal electrons
increases and the plasma becomes thermally dominated. Together with
the resulting break in the distribution of the non-thermal electrons
this results in a cutoff of the photon spectrum.

The rate at which non-thermal electrons appear in the hybrid electron
distribution can be written as a power law (Gierli\'nski \etal 1999)

\begin{equation}
\dot N_{nt} = \frac{d N_{nt}(\gamma)}{d t}\propto \gamma^{-\Gamma_{in}}.
\end{equation}

\noindent Here, $\gamma$ is the Lorentz factor of the non-thermal
electrons and $\Gamma_{in}$ the power law slope of the injected soft
photon distribution, which influences the slope of the power law
spectrum as

\begin{equation}
\Gamma_{in} \simeq 2(\Gamma -1). 
\end{equation}

The observed power law slope $\Gamma \sim$2.5 ($\Gamma \sim$3.5)
implies $\dot N_{nt}\propto\gamma^{-3}$ ($\propto
\gamma^{-5}$). Therefore the fraction of non-thermal electrons
decreases with steepening power law.

In \grs, the observed continuous distribution of $\Gamma$ between 2.4
and 3.5 is evidence for a variability in the power law component and
thus in the composition and locus of the electron distribution. From
the observation of the reflection and the 0.5--10\,Hz QPO it is
conspicuous that the electron distribution must have a small (relative
to the system) spatial size. As a small compactness, e. g. large
plasma volume, results in a spectral cutoff, for the $\chi$-states of
\grs\ a hybrid electron distribution not can be ruled out. It is
difficult to distinguish spectroscopically between a hybrid and a
thermal plasma if the hybrid plasma is thermally dominated. The only
possibility is the search for the predicted photon excess at high
energies and for the annihilation line (Coppi 1999). For a better
understanding, enhanced high energy detectors, such as on INTEGRAL,
are required. The analysis of such spectra also needs further
developed physical models, instead of a simple power law.

\bigskip

The investigation of the {\it RXTE} data of \grs\ with the
DISKBB+REFSCH model does not allow a definite conclusion about the
origin of the hard spectral component. Only the BMC model is ruled out
for the $\chi$-states. Though a clear distinction between a thermal
and a hybrid not can be proposed, it seems clear that the thermal
electrons dominate.

It has been shown above (Fig.~\ref{alpK}) that two different states of
comptonization with two different pivoting energies are observed. Very
recently (Zdziarski \etal 2002) pivoting was found also in \cyg. A
varying amount of soft seed photons undergoing the comptonization
process probably accounts for the pivoting of the spectrum. One
possible explanation is a constant disk blackbody and an
expanding/contracting hot coronal plasma. A change in the size of the
corona leads to a change in the fraction of intercepted soft photons.
Assuming constant optical plasma depth, $\tau_T$, if the soft
luminosity is high, the electrons in the corona are cooled efficiently
and the spectrum of the comptonized photons is soft. On the other
hand, if the soft luminosity decreases, the temperature of the corona
increases and the spectrum hardens). The ocurrance of two differing
pivoting branches suggests the existence of two states with different
$\tau_T$ and/or different electron composition (thermal/non-thermal).

\subsection{Reflection}

The reflection of X-rays on an accretion disk manifests itself in two
specific spectral features: (i) the characteristic emission line
spectrum (mainly the K$\alpha$ lines of the most common metals) with
the 6.4\,keV iron emission lines as the strongest and (ii) a
characteristic hump arises above 10\,keV because of the energy
dependence of the cross section of absorption and Compton scattering.

Observations with {\it ASCA} in different variability states indicate
a variable iron absorption or emission in \grs\ (Ebisawa \etal
1997). The iron emission at 6.4\,keV dominates the spectra from
25.10.1996 and 25.04.1997 whereas a distinct absorption at 7\,keV and
no emission was seen on 27.09.1994 and 26.04.1995. During the
25.04.1997 observation \grs\ switched between $\chi$- and
$\alpha$-states (based on {\it RXTE} data). Therefore, it is not clear
whether the emission line originates during the $\chi$- or during the
$\alpha$-interval. No statement can be made about the states during
the other {\it ASCA} observations due to lack of corresponding {\it
RXTE} observations.

Recent {\it Chandra} data of \grs\ obtained in the low hard state
revealed neutral K absorption edges, ionized resonance absorption from
Fe (XXV, XXVI) and possible emission from neutral Fe K$\alpha$ and
ionized Fe XXV (Lee \etal 2002), suggesting conditions favorable for
reflection.

\subsubsection{Reflection in LMXB and AGN}

Investigations of Seyfert galaxies and X-ray binaries in the low/hard
state show a correlation between the slope of the hard X-ray component
and the reflection. Incoming radiation with a steeper slope is more
reflected than radiation with a flatter slope.

The reflection is stronger in black hole candidates than in Seyfert
galaxies for a given $\Gamma$ (Zdziarski 1999). This can be explained
using an optical thick accretion disk. The different masses of the
black holes imply different maximum energies of the blackbody photons
(Svensson 1996). The hard X-ray photons from the corona can ionize the
upper layers of the accretion disk and an ionized layer above neutral
matter evolves. If $\Gamma$ is large the heating of the disk is small
and the cold layer of the disk lies near the disk surface. With
decreasing power law slope the temperature of the upper layer and
therefore the ionization increases, the absorption is smaller, more
photons are scattered and the characteristic reflection hump is
suppressed, sufficient to explain the observed
$R$($\Gamma$)-correlation.

The correlation in Seyfert galaxies is generally steeper than that
observed in X-ray binaries (Zdziarski 1999). In Seyferts the
correlation has been explained with a model consisting of a static,
thermal corona above a neutral reflector (Svensson 1996). The
existence of the R($\Gamma$)-correlation implies a feedback where the
existence of the reflecting matter influences the hardness of the
X-ray spectrum (B\"ottcher \etal 1998, Zdziarski \etal 1999). Assuming
that the cold matter (the accretion disk) emits soft photons which
become seed photons of the comptonization, than with increasing solid
angle of the reflector (here the accretion disk) the flux of soft
photons increases and the cooling rate of the hot corona
increases. For a thermal plasma the resulting power law component
steepens with increasing cooling rate.

On the other hand, models which are based on non-thermal electrons
should not show a dependence of spectral hardness and cooling rate
(Lightman \& Zdziarski 1987, Zdziarski \etal 1999). Models in which
the seed photons are intrinsically produced in the hot plasma
(e. g. synchrotron radiation) not can reproduce the observed
correlation. The power law slope is then independent of the
reflection.

\subsubsection{Reflection in GRS 1915+105}

The $\chi$-states in \grs\ show a similiar correlation as seen in
Seyfert galaxies and X-ray binaries (Zdziarski \etal 1999). The
reflection amplitude is strongly effected by the disk model. The
higher the disk temperature, the larger the reflection amplitude. The
disk temperatures in X-ray binaries are higher than in Seyfert
galaxies. In \grs\ the temperature of the disk is still higher,
explaining why a majority of the \grs\ data lies above the correlation
function (showing slightly higher $R$ for a given $\Gamma$) in
Fig.~\ref{bhc-corr}.

The reflection amplitude of $R> 1$ strongly suggests an anisotrophic
inverse Compton process. The dominant fraction of the up-scattered
soft disk photons is directed back into the plane of the accretion
disk. This produces the observed large reflection amplitude in
contrast to an isotrophic scattering where $R\leq 1$.

Observations of \grs\ show a strongly varying $\Gamma$, therefore the
corona can hardly be static above the reflector. In \grs, $\Gamma$
increases with increasing radio emission. Higher radio emission may
imply more outflow away from the disk with $\beta=(v/c)>$0. Thus, the
higher the mass outflow, the lower the electron temperature which is
required to produce the observed amount of comptonization. On the
other hand, with a dynamical, thermal corona the spectrum hardens with
increasing $\beta$ due to relativistic aberration, giving a flatter
power law with higher outflow velocity (Malzac \etal 2001).

The observation in \grs\ predicts an at least partly thermal source of
the hard X-ray photons and an important contribution of the cooling
due to the soft photons. This allows both a thermal and a hybrid
electron distribution.

It is more difficult to interpret the behavior of observations with
high $\Gamma$ and small $R$ in Fig.~\ref{bhc-corr}. Unlike in a
thermal plasma, the Compton scattering in a non-thermal plasma does
not depend on the cooling rate of the soft photons, instead it depends
on the slope of the electron distribution (Poutanen \& Coppi 1998). A
change in the properties of the plasma can provide the differing
behavior.

For a better determination of the correlation and for conclusions
about the distinct system components, more complex ionization models
are needed. They partly exist (Done \& Nayakshin 2000) but require too
much computing power to be useful for the analysis of real spectra at
this stage.

\subsection{X-ray-Radio-Correlation}

The lack of a correlation of the radio flux with any of the disk
parameters is somewhat unexpected since the (short-duration) low/hard
states have earlier been related to jet ejections with radio emission
by synchrotron radiation of the ejected plasma (Pooley \& Fender
1997). Since IR (Eikenberry \etal 2000) and radio observations (Pooley
\& Fender 1997, Fender \& Pooley 2000) have led to the conclusion that
\grs\ shows jets on various scales, or even has a continuous
distribution of jet strength, one would have expected that also in
$\chi$-states the radio flux correlates with the disk
temperature/emissivity.

The matter in the sporadic, relativistic jets may originate from
disruption of the inner part of the accretion disk during
$\beta$-states (Mirabel \etal 1998). In $\chi$-states no correlation
of the radio emission and the inner disk radius is observed. Instead,
the hard spectral component is correlated with the radio flux.

Muno \etal (1999) found a positive correlation of the accretion disk
temperature and the frequency of the 0.5--10\,Hz quasi periodic
oscillations, and of the radio flux with the QPO frequency (Muno \etal
2001). With increasing $T_{bb}$ and decreasing radio emission the QPO
frequency increases. This predicts a negative correlation of $T_{bb}$
and radio flux. The fact that nothing like this is found in the
present analysis can have several reasons. As mentioned before, Muno
\etal (1999) used the standard model DISKBB+BKNPO for the analysis of
the {\it RXTE} spectra. This seems not suitable for the $\chi$-states
of \grs. Accordingly, the disk temperature and therefore the
correlation of the QPO frequency and the disk temperature have to be
interpreted carefully.

It is generally assumed that both the radio jets (Fendt \& Greiner
2001, Fender 2001) and the hard spectral component originate near the
black hole. With increasing outflow of matter (therefore increasing
radio emission) the interaction with the electron distribution becomes
significant.  Similar to the interpretation of the pivoting behavior
of the X-ray spectra, an increasing outflow of matter implies an
increasing size of the scattering medium and therefore an increasing
amount of intercepted soft seed photons. This leads to a lower plasma
temperature due to cooling and to a softer X-ray spectrum, which is
seen in the $\Gamma$($F_R$)-correlation. Simultaneously, the
outflowing matter intermingles with the coronal matter and pushes it
away from the accretion disk. The formerly thermally dominated
electron distribution may become non-thermal dominated. This should
result in a shift of the cutoff of the power law component in the
$\chi$-states with higher radio emission above the HEXTE range to
higher energies.

When the radio emission is low, the observed power law is supposed to
originate in the corona above the accretion disk. Assuming that the
radio quiet state lacks outflow of matter or the effect of the outflow
on the corona is negligible, the hard spectral component possibly
originates due to comptonization of disk photons in the corona. A
thermal electron distribution can then explain the X-ray spectra of
the radio quiet $\chi$-states.

The hard X-ray component in $\chi$-states therefore comes most likely
from soft disk photons which are inversely scattered on a thermal
dominated electron distribution (when radio flux is low) or on the
base if a continuous jet (when radio flux increases). The suggestion
of the base of the jet as source of the hard X-ray photons was already
made by Fender (2001). Only the increasing reflection with increasing
radio flux is still unresolved. 

In conflict with the anti-correlation of $F_X$(20--200) and $F_R$ in
\grs\ a linear correlation is observed on the LMXB \gx\ (Corbel \etal
2000). With increasing radio flux the X-ray flux increases. This is
true for the hard (20--100\,keV) and for the soft X-ray flux
(2--12\,keV) where \grs\ shows no correlation. The soft disk component
in \gx\ is negligible above 2\,keV and the power law component alone
quantifies the flux. Although \grs\ shows an equal
$R$($\Gamma$)-correlation as \gx\ the $\Gamma$($F_R$)- and
$F_X$($F_R$)-correlation is opposite. That is remarkable because the
$R$($\Gamma$)-correlation suggests a similar structure near the black
hole (especially the corona) in \grs\ and \gx. No such correlation of
$F_X$ and $F_R$ is observed in \cyg\ (Brocksopp \etal 1999).

In principle the hard X-ray photons can originate in the continuous
jet itself. But following the equations of Marscher (1983) the hard
X-ray photons from self comptonization of a compact syncrotron jet are
negligible with regard to the coronal contribution.  An analogous
result was found for \cyg\ and \gx. The contribution of thermal X-ray
bremsstrahlung in the jet is also too small to explain the observed
hard X-ray luminosities (Memola \etal 2002).

\section{Conclusion}

The analysis of 139 {\it RXTE} observations of \grs\ in the
$\chi$-state revealed variable components and parameters of the X-ray
spectra. The structure of the hard X-ray power law depends e. g. on
the radio flux. Further, a two-branch correlation of the power law
slope and the power law normalization was found. Theses branches show
different pivoting behavior.

The most probable geometry is that of a hot corona above the accretion
disk. A bulk motion comptonization seems to be ruled out for the
$\chi$-states because of the time lags of hard and soft X-ray photons.

The continuous outflow of matter is not correlated with the accretion
disk parameters as measured at X-rays. Neither the disk temperature
nor the inner disk radius are connected with the radio
flux. Therefore, the outflowing matter should not be provided by a
disruption of the inner part of the accretion disk, as thought to be
the case for the sporadic jets in \grs, and a positive correlation
between radio emission and power law slope is found. Because both the
base of the jet and the corona are probably located near the black
hole the correlation indicates an interaction of both
structures. Whereas during low radio emission the hard X-ray component
originates due to comptonization on a thermally-dominated corona, in
radio loud states the comptonization should appear in the outflowing
matter.

\grs\ shows a positive $R$($\Gamma$)-correlation as seen in other
X-ray binaries and AGN. The large reflection amplitude suggests a
highly anisotrophic inverse Compton scattering with the dominant part
of the soft photons being scattered back into the disk plane.

\begin{acknowledgements}

The authors thank E. H. Morgan (MIT) and G. G. Pooley (MRAO) for
providing the ASM and {\it RT} data and W. A. Heindl ({\it RXTE}) and
K. A. Arnaud (XSPEC) for their comments. We thank the referee,
A. Zdziarski, for the thorough reading and several suggestions which
improved the presentation and T. Belloni for the constructive
criticism. The {\it Ryle Telescope} is supported by PPARC. The {\it
Green Bank Interferometer} is a facility of the National Science
Foundation operated by the NRAO in support of NASA High Energy
Astrophysics programs.

\end{acknowledgements}

\end{document}